\documentclass[conference]{IEEEtran}
\IEEEoverridecommandlockouts
\usepackage{tikz}
\usetikzlibrary{calc,positioning}
\nonstopmode
\usepackage{cite}
\usepackage{amsmath,amssymb,amsfonts}
\usepackage[linesnumbered,lined,ruled,vlined]{algorithm2e}
\usepackage{graphicx}
\usepackage{textcomp}
\usepackage{xcolor}
\usepackage[bookmarks=false]{hyperref}
\usepackage{cleveref}
\usepackage{commath}
\usepackage{optidef}
\usepackage[symbol]{footmisc}
\usepackage{lipsum}
\usepackage{enumitem}
\usepackage{booktabs}

\allowdisplaybreaks
\usepackage[left=1.62cm,right=1.62cm,top=1.7cm,bottom=2.67cm]{geometry}
\IEEEaftertitletext{\vspace{-0.5\baselineskip}}

\makeatletter
\newcommand{\nosemic}{\renewcommand{\@endalgocfline}{\relax}}
\newcommand{\dosemic}{\renewcommand{\@endalgocfline}{\algocf@endline}}

\let\oldnl\nl
\newcommand{\nonl}{\renewcommand{\nl}{\let\nl\oldnl}}
\makeatother

\renewcommand{\arraystretch}{1.25}

\definecolor{cardinal}{HTML}{8C1515}

\def\BibTeX{{\rm B\kern-.05em{\sc i\kern-.025em b}\kern-.08em
    T\kern-.1667em\lower.7ex\hbox{E}\kern-.125emX}}

\DeclareMathOperator*{\maximize}{maximize}
\DeclareMathOperator*{\minimize}{minimize}

\begin{document}

\title{Canonical Optimization for MIMO MAC Design}
\author{
    \IEEEauthorblockN{
        Muhammad Umer\IEEEauthorrefmark{1}, Muhammad Ahmed Mohsin\IEEEauthorrefmark{1}, Ahsan Bilal\IEEEauthorrefmark{2}, and John M. Cioffi\IEEEauthorrefmark{1}}
    \IEEEauthorblockA{
        \IEEEauthorrefmark{1}Department of Electrical Engineering, Stanford University, Stanford, CA 94305, USA\\
        \IEEEauthorrefmark{2}School of Computer Science, University of Oklahoma, Norman, OK 73019, USA}
    \IEEEauthorblockA{Email: \{mumer, muahmed, cioffi\}@stanford.edu, ahsan.bilal-1@ou.edu}
}
\maketitle
\begin{abstract}
    Resource allocation in the multiple-input multiple-output (MIMO) multiple access channel (MAC) is a fundamental problem in multiuser communications, yet it is increasingly treated as non-convex and computationally intractable. This has motivated a large body of heuristic machine learning and successive-approximation methods. Results here show that the MIMO MAC admits canonical convex formulations and present four solvers that together characterize its capacity region. \textsc{maxRMAC} performs weighted sum-rate maximization under per-user energy constraints, \textsc{minPMAC} finds the minimum weighted energy required to support target rates, \textsc{maxRESMAC} performs weighted sum-rate maximization under a total energy constraint, and \textsc{admMAC} tests rate-region feasibility. The solvers exploit the polymatroid structure of the MAC rate region and the separability of the dual Lagrangian across frequency tones, which reduces the problem to parallel per-tone covariance optimizations solved via limited-memory Broyden--Fletcher--Goldfarb--Shanno (L-BFGS) over Cholesky-like covariance factors. Experiments on spatially correlated MIMO orthogonal frequency-division multiplexing (OFDM) channels show that the proposed solvers match a commercial convex solver in solution quality while running up to two orders of magnitude faster and scaling to regimes where the commercial solver times out. Through broadcast channel (BC) to MAC duality, the same solvers also enable optimal precoder design for the MIMO BC. All solvers are open-sourced and available at \href{https://github.com/muhd-umer/canonical-mac}{\textcolor{cardinal}{\texttt{https://github.com/muhd-umer/canonical-mac}}}.
\end{abstract}
\section{Introduction}\label{sec:intro}
Resource allocation in the multiple-input multiple-output (MIMO) multiple access channel (MAC) poses three fundamental questions for any set of uncoordinated transmitters sharing a common receiver: \textit{what rates can the users simultaneously achieve, how should their transmit covariances be chosen to maximize throughput or minimize power, and is a given operating point feasible at all?} These questions arise in virtually every transmission system, from cellular base stations to wireline vectored digital subscriber lines.

These problems' information-theoretic foundations are well established. The MAC capacity region for Gaussian channels was characterized through mutual-information chain rules and the polymatroid structure of achievable rate vectors~\cite{737513}. Iterative waterfilling was shown to achieve the sum-capacity of the vector MAC under sum-power constraints~\cite{1262622}, and weighted sum-rate maximization together with weighted sum-power minimization further characterized the MIMO MAC capacity region under individual power constraints~\cite{1665015}, with the two problems linked via Lagrangian duality. Broadcast channel (BC) to MAC duality, which enables BC design through an equivalent dual MAC, was established in~\cite{1291726} and extended to the MIMO setting with dirty-paper coding in~\cite{1327794}.

Despite this well-understood convex structure, multiuser MIMO resource allocation is increasingly treated as non-convex and computationally intractable. A growing body of work applies deep reinforcement learning, deep neural networks, and other machine learning methods to MAC and non-orthogonal multiple access (NOMA) power allocation and user scheduling~\cite{wang2024spectrum,mohsan2023survey}, often citing non-convex formulations as motivation. Although data-driven approaches offer speed advantages for real-time inference, they sacrifice optimality guarantees and provide no structural insight into the solution. Weighted minimum mean square error (WMMSE)-based algorithms~\cite{5199574} and successive convex approximation (SCA) methods~\cite{1665016} have also been widely adopted; however, they do not exploit the full polymatroid structure of the MAC rate region and may converge to local optima for the general MIMO MAC with per-user constraints. The admission problem, that is, determining whether a target rate vector is feasible, was formulated in~\cite{1665015} but has received comparatively little algorithmic attention. What has been missing is not theory but practice, namely scalable, open solvers that realize the canonical convex formulation with global optimality guarantees.

This paper fills that gap with four canonical solvers for MIMO MAC resource allocation, each convex, globally optimal, and scalable. All four exploit the polymatroid structure of the MAC rate region and its associated Lagrangian decomposition across frequency tones, reducing each problem to parallel per-tone subproblems solved via limited-memory Broyden--Fletcher--Goldfarb--Shanno (L-BFGS)~\cite{liu1989limited} over Cholesky-like factors of the transmit covariance matrices. These implicitly enforce the positive semidefinite (PSD) constraint. The outer dual optimization uses an ellipsoid subgradient algorithm for per-user energy constraints, or bisection for a single sum-energy constraint. Both enjoy polynomial convergence-time guarantees. The four solvers are:
\begin{enumerate}[leftmargin=*]
    \item \textsc{maxRMAC}---weighted sum-rate maximization under per-user energy constraints;
    \item \textsc{minPMAC}---minimum weighted energy for target rates, with automatic decoding-order selection and time-sharing;
    \item \textsc{maxRESMAC}---weighted sum-rate maximization under a total energy constraint; and
    \item \textsc{admMAC}---feasibility testing for a given rate-energy pair, determining membership in the MAC capacity region.
\end{enumerate}
Together, these solvers completely characterize the MAC capacity region and, through BC-MAC duality~\cite{1291726}, also enable optimal BC precoder design.


\section{Preliminaries and System Model}\label{sec:system_model}
\begin{figure}[t]
    \centering
    \includegraphics[width=0.985\linewidth]{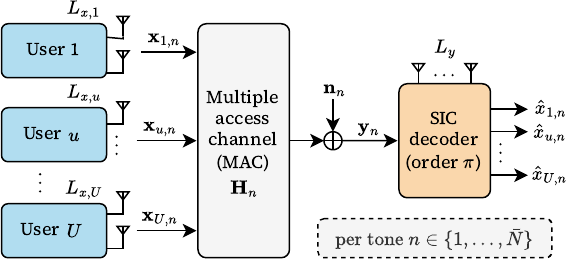}
    \caption{\textbf{MIMO MAC system model.} $U$ transmitters with per-user antenna counts $L_{x,u}$ communicate to a single receiver with $L_y$ antennas across $\bar{N}$ frequency tones. The receiver applies SIC in an order determined by the rate weights~$\boldsymbol{\theta}$.}
    \label{fig:mac_system}
\end{figure}


\subsection{Channel model}\label{sec:channel}

Consider a $U$-user MIMO MAC with $L_y$ receive antennas and $\bar{N}$ orthogonal frequency tones, as shown in Fig.~\ref{fig:mac_system}. User~$u$ has $L_{x,u}$ transmit antennas. On tone~$n$, the received signal is
\begin{equation}\label{eq:mac}
    \mathbf{y}_n = \sum_{u=1}^{U} \mathbf{H}_{u,n}\,\mathbf{x}_{u,n} + \mathbf{n}_n,
\end{equation}
where $\mathbf{H}_{u,n}\!\in\!\mathbb{C}^{L_y\times L_{x,u}}$ is the noise-whitened channel matrix for user~$u$ on tone~$n$, $\mathbf{x}_{u,n}$ is user~$u$'s transmitted signal with covariance $\mathbf{R}_{u,n}\!=\!\mathbb{E}[\mathbf{x}_{u,n}\mathbf{x}_{u,n}^*]\!\succeq\!0$, and $\mathbf{n}_n\!\sim\!\mathcal{CN}(\mathbf{0},\mathbf{I})$. The aggregate channel is $\mathbf{H}_n = [\mathbf{H}_{1,n}\;\cdots\;\mathbf{H}_{U,n}] \in \mathbb{C}^{L_y\times L_{T}}$, where $L_{T}=\sum_u L_{x,u}$. Each user's total transmitted energy and rate are $\mathcal{E}_u = \sum_{n}\mathrm{tr}(\mathbf{R}_{u,n})$ and $b_u = \sum_{n} b_{u,n}$, respectively.

\subsection{MAC capacity region and polymatroid structure}\label{sec:rate_region}
The receiver decodes users via successive interference cancellation (SIC). Let $\mathbf{Q}_{u,n} \triangleq \mathbf{H}_{u,n}\mathbf{R}_{u,n}\mathbf{H}_{u,n}^*$ denote user~$u$'s received signal covariance on tone~$n$. Under decoding order $\boldsymbol{\pi}$, the user at position~$k = \boldsymbol{\pi}^{-1}(u)$ achieves the rate
\begin{equation}\label{eq:sic_rate}
    b_{u,n} = \frac{1}{c_b}\log_2\frac{\big|\mathbf{I}+\sum_{j=k}^{U}\mathbf{Q}_{\pi(j),n}\big|}{\big|\mathbf{I}+\sum_{j=k+1}^{U}\mathbf{Q}_{\pi(j),n}\big|},
\end{equation}
where $c_b\!=\!1$ for complex and $c_b\!=\!2$ for real baseband. For fixed covariances, the achievable rate vectors across all $U!$ orderings form a \emph{polymatroid} characterized by the subset-sum constraints
\begin{equation}\label{eq:polymatroid}
    \sum_{u\in\mathcal{S}} b_{u,n} \;\leq\; \frac{1}{c_b}\log_2\!\Big|\mathbf{I}+\sum_{u\in\mathcal{S}}\mathbf{Q}_{u,n}\Big|, \quad \forall\,\mathcal{S}\!\subseteq\!\{1,\ldots,U\},
\end{equation}
where each decoding order corresponds to a vertex of this polymatroid. The MAC capacity region $\mathcal{C}(\boldsymbol{\mathcal{E}})$ under per-user energy constraints $\boldsymbol{\mathcal{E}}=(\mathcal{E}_1,\ldots,\mathcal{E}_U)^T$ is the convex hull of all such vertices over all valid covariance choices. This polymatroid structure is central to the solvers, as it determines the optimal SIC ordering for any weight vector and governs when time-sharing across orderings is required.

\subsection{Building blocks for the solvers}\label{sec:building_blocks}

Section~\ref{sec:solvers}'s four solvers share the following two key computational building blocks;

\subsubsection{Lagrangian decomposition across tones}\label{sec:lagrangian}
A structural property exploited here is that the Lagrangian of every MAC optimization problem \emph{decomposes across tones}, reducing a large joint problem to $\bar{N}$ independent per-tone subproblems. To see this, let $\boldsymbol{\theta}\!=\!(\theta_1,\ldots,\theta_U)^T\!\geq\!\mathbf{0}$ be rate weights with sorted order $\theta_{(1)}\!\geq\!\cdots\!\geq\!\theta_{(U)}$, and define the weight differences
\begin{equation}\label{eq:delta}
    \delta_k = \theta_{(k)} - \theta_{(k+1)},\quad k=1,\ldots,U,\quad \theta_{(U+1)}\triangleq 0.
\end{equation}
Applying summation by parts yields $\sum_u\theta_u b_{u,n} = \frac{1}{c_b}\sum_{k=1}^U\delta_k\log_2|\mathbf{S}_{k,n}|$, where $\mathbf{S}_{k,n} = \mathbf{I} + \sum_{j=1}^{k}\mathbf{Q}_{(j),n}$ is the cumulative signal-plus-noise covariance after including the $k$ highest-weight users and $\mathbf{Q}_{(j),n}$ denotes the received covariance of the user with the $j$-th largest $\theta$. This expression corresponds to decoding users from lowest $\theta$ to highest $\theta$, so that the highest-weight user is decoded last and sees no residual interference. This is the ordering that maximizes $\sum_u\theta_u b_u$ over all $U!$ permutations. Introducing energy multipliers $\mathbf{w}\!\geq\!\mathbf{0}$, the Lagrangian 
separates as $\mathcal{L} = \sum_n \ell_n + \mathbf{w}^{\!\top}\boldsymbol{\mathcal{E}}$, 
with the \emph{per-tone Lagrangian}
\begin{equation}\label{eq:tone_lag}
    \ell_n = \frac{1}{c_b}\sum_{k=1}^{U}\delta_k\log_2|\mathbf{S}_{k,n}| - \sum_{u=1}^{U}w_u\,\mathrm{tr}\big(\mathbf{R}_{u,n}\big).
\end{equation}
Since tones couple only through the outer dual variables $\mathbf{w}$, all $\bar{N}$ per-tone subproblems can be parallelized across available hardware.

\subsubsection{L-BFGS over Cholesky-like factors}\label{sec:lbfgs}
Each per-tone subproblem maximizes~\eqref{eq:tone_lag} over $\{\mathbf{R}_{u,n}\!\succeq\!0\}_{u=1}^U$. Parameterizing each covariance as $\mathbf{R}_{u,n}=\mathbf{B}_u\mathbf{B}_u^*$ via a Cholesky-like factor $\mathbf{B}_u\!\in\!\mathbb{C}^{L_{x,u}\times L_{x,u}}$ converts the PSD-constrained problem into an unconstrained one. The real and imaginary parts of all $\mathbf{B}_u$ blocks are packed into a single real vector $\mathbf{z}\!\in\!\mathbb{R}^{d}$, $d\!=\!2\sum_u L_{x,u}^2$, and L-BFGS with Armijo line search~\cite{liu1989limited} is applied. Warm-starting from the previous outer-loop iterate provides substantial acceleration in practice.

\begin{figure}[t]
    \centering
\includegraphics[width=0.9\linewidth]{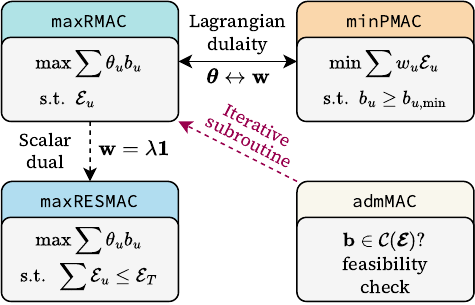}
    \caption{\textbf{Relationships among the four canonical MAC solvers.} \textsc{maxRMAC} and \textsc{minPMAC} are Lagrangian duals, where rate weights $\boldsymbol{\theta}$ in \textsc{maxRMAC} become dual variables in \textsc{minPMAC} and energy multipliers $\mathbf{w}$ swap roles correspondingly. \textsc{maxRESMAC} is a scalar-constraint specialization of \textsc{maxRMAC}. \textsc{admMAC} calls \textsc{maxRMAC} as a subroutine in an outer feasibility loop.}
    \label{fig:solver_relationships}
\end{figure}

\section{Canonical MAC Solvers}\label{sec:solvers}
This section presents four solvers that together address all fundamental MAC resource allocation problems. Each builds on the tone decomposition and L-BFGS inner solver, differing only in the outer-loop structure dictated by the constraint type. Their relationships are illustrated in Fig.~\ref{fig:solver_relationships}, and Table~\ref{tab:solvers} provides a summary.

\begin{table}[h]
    \centering
    \caption{Summary of canonical MAC solvers.}
    \label{tab:solvers}
    \renewcommand{\arraystretch}{1.15}
    \renewcommand{\tabcolsep}{14pt}
    \footnotesize
    \begin{tabular}{@{}lll@{}}
        \toprule
        \textbf{Solver}    & \textbf{Problem}                             & \textbf{Outer method}                  \\
        \midrule
        \textsc{maxRMAC}   & Max weighted rate sum,                       & Ellipsoid on                           \\
                           & per-user energy $\boldsymbol{\mathcal{E}}$   & $\mathbf{w}\in\mathbb{R}^U_+$          \\[1pt]
        \textsc{minPMAC}   & Min weighted energy,                         & Ellipsoid on                           \\
                           & target rates $\mathbf{b}_{\min}$             & $\boldsymbol{\theta}\in\mathbb{R}^U_+$ \\[1pt]
        \textsc{maxRESMAC} & Max weighted rate sum,                       & Bisection on                           \\
                           & total energy $\mathcal{E}_{T}$               & $\lambda\in\mathbb{R}_+$               \\[1pt]
        \textsc{admMAC}    & Feasibility of                               & Adaptive $\boldsymbol{\theta}$         \\
                           & $(\mathbf{b},\boldsymbol{\mathcal{E}})$ pair & + Frank-Wolfe                          \\
        \bottomrule
    \end{tabular}
\end{table}

\subsection{{\upshape\textsc{maxRMAC}}: Weighted sum-rate maximization}\label{sec:maxrmac}
\subsubsection{Problem formulation}
Given per-user energy budgets $\boldsymbol{\mathcal{E}}=(\mathcal{E}_1,\ldots,\mathcal{E}_U)^T$ and rate weights $\boldsymbol{\theta}\!\geq\!\mathbf{0}$,
\begin{equation}\label{eq:maxrmac}
    \begin{aligned}
        \maximize_{\{\mathbf{R}_{u,n}\succeq 0\}} & \;\; \sum_{u=1}^{U}\theta_u\, b_u                                                             \\
        \text{s.t.}                               & \;\; \sum_{n=1}^{\bar{N}}\mathrm{tr}\big(\mathbf{R}_{u,n}\big)\leq\mathcal{E}_u,\;\forall\,u.
    \end{aligned}
\end{equation}
Sweeping $\boldsymbol{\theta}$ over the positive orthant traces the entire boundary of the capacity region $\mathcal{C}(\boldsymbol{\mathcal{E}})$.

\subsubsection{Solution via ellipsoid method}
Dualizing the energy constraints introduces per-user multipliers $\mathbf{w}\!\geq\!\mathbf{0}$, and the Lagrangian decomposes as $\mathcal{L}=\sum_n \ell_n + \mathbf{w}^T\boldsymbol{\mathcal{E}}$, where $\ell_n$ is given in~\eqref{eq:tone_lag}. Each tone is solved independently via L-BFGS, parallelized across all $\bar{N}$~tones.

The outer optimization searches over $\mathbf{w}$ using the \emph{ellipsoid method}. At each iteration, the subgradient of the dual function with respect to~$\mathbf{w}$ is
\begin{equation}\label{eq:subgrad_maxr}
    g_u = \mathcal{E}_u - \sum_{n=1}^{\bar{N}}\mathrm{tr}\big(\hat{\mathbf{R}}_{u,n}\big),
\end{equation}
where $\hat{\mathbf{R}}_{u,n}$ denotes the per-tone optimizer for the current~$\mathbf{w}$. The ellipsoid update rule is
\begin{equation}\label{eq:ellipsoid}
    \begin{aligned}
        \tilde{\mathbf{g}} & = \mathbf{g}/\sqrt{\mathbf{g}^T\mathbf{A}\mathbf{g}}, \quad
        \mathbf{w} \leftarrow \mathbf{w} - \tfrac{1}{U+1}\mathbf{A}\tilde{\mathbf{g}},                                                           \\
        \mathbf{A}         & \leftarrow \tfrac{U^2}{U^2-1}\big(\mathbf{A} - \tfrac{2}{U+1}\mathbf{A}\tilde{\mathbf{g}}\tilde{\mathbf{g}}^T\mathbf{A}\big),
    \end{aligned}
\end{equation}
where $\mathbf{A}$ is the ellipsoid shape matrix and non-negativity $\mathbf{w}\!\geq\!\mathbf{0}$ is enforced by cutting-plane projections whenever a component becomes negative. After convergence, the covariances are rescaled as $\mathbf{R}_{u,n}\leftarrow(\mathcal{E}_u/\hat{\mathcal{E}}_u)\,\mathbf{R}_{u,n}$, where $\hat{\mathcal{E}}_u\!=\!\sum_n\mathrm{tr}(\hat{\mathbf{R}}_{u,n})$, which ensures exact constraint satisfaction. Algorithm~\ref{alg:maxrmac} summarizes the procedure.

\begin{algorithm}[h]
    \DontPrintSemicolon
    \KwIn{$\mathbf{H}$, $\boldsymbol{\mathcal{E}}$, $\boldsymbol{\theta}$, $c_b$}
    \KwOut{$\{\mathbf{R}_{u,n}\}$, $\mathbf{w}$, $\{b_{u,n}\}$}
    Initialize ellipsoid $(\mathbf{A},\mathbf{w})$\;
    \For{$\mathrm{iter}=1,2,\ldots$}{
        \ForPar{$n=1,\ldots,\bar{N}$}{
            Solve per-tone Lagrangian~\eqref{eq:tone_lag} via L-BFGS\;
        }
        $g_u \leftarrow \mathcal{E}_u - \sum_n\mathrm{tr}(\hat{\mathbf{R}}_{u,n})$ for all $u$\;
        \lIf{$\sqrt{\mathbf{g}^T\mathbf{A}\mathbf{g}}\leq\epsilon$}{\textbf{break}}
        Update $(\mathbf{w},\mathbf{A})$ via~\eqref{eq:ellipsoid}; project $\mathbf{w}\geq\mathbf{0}$\;
    }
    Rescale $\mathbf{R}_{u,n}\leftarrow(\mathcal{E}_u/\hat{\mathcal{E}}_u)\,\mathbf{R}_{u,n}$\;
    Compute rates $\{b_{u,n}\}$ via~\eqref{eq:sic_rate}\;
    \caption{\textsc{maxRMAC}}\label{alg:maxrmac}
\end{algorithm}

\subsection{{\upshape\textsc{minPMAC}}: Minimum power for target rates}\label{sec:minpmac}
\subsubsection{Problem formulation}
The dual perspective asks a complementary question. Given target rates $\mathbf{b}_{\min}\!=\!(b_{1,\min},\ldots,b_{U,\min})^T$ and energy weights $\mathbf{w}\!>\!\mathbf{0}$, what is the minimum weighted energy needed to support them?
\begin{equation}\label{eq:minpmac}
    \begin{aligned}
        \minimize_{\{\mathbf{R}_{u,n}\succeq 0\}} & \;\; \sum_{u=1}^{U} w_u\,\mathcal{E}_u \\
        \text{s.t.}                               & \;\; b_u \geq b_{u,\min},\;\forall\,u.
    \end{aligned}
\end{equation}
This problem is the Lagrangian dual of~\eqref{eq:maxrmac}. The roles of $\boldsymbol{\theta}$ and $\mathbf{w}$ interchange, with $\boldsymbol{\theta}\!\geq\!\mathbf{0}$ now serving as dual variables for the rate constraints. Under Slater's condition~\cite{boyd2004convex}, strong duality holds and the duality gap is zero.

\subsubsection{Solution via ellipsoid method with time-sharing}
The ellipsoid method now searches over $\boldsymbol{\theta}\!\geq\!\mathbf{0}$ with subgradient $g_u = b_u^\star - b_{u,\min}$, where $b_u^\star\!=\!\sum_n b_{u,n}^\star$ is the total achieved rate for user~$u$ at the current~$\boldsymbol{\theta}$. Since the \textsc{minPMAC} dual is a maximization over $\boldsymbol{\theta}$, the true subgradient of the dual objective is $\mathbf{b}_{\min}-\mathbf{b}^\star = -\mathbf{g}$, so applying the ellipsoid step~\eqref{eq:ellipsoid} in the $-\tilde{\mathbf{g}}$ direction correctly implements gradient ascent on the dual.

At convergence, the optimal $\boldsymbol{\theta}$ gives the decoding-order structure through \emph{clustering}. Users with equal $\theta_u$ values form macro users, and any permutation within a cluster yields an equivalent polymatroid vertex. The inter-cluster ordering is fixed by the relative cluster $\theta$ values, reducing the number of relevant orderings from $U!$ to $\prod_c |\mathcal{S}_c|!$, where $\mathcal{S}_c$ is the $c$-th cluster. When no single ordering meets all rate targets, the solver recovers \emph{time-sharing} fractions $\{\alpha_k\!\geq\!0\}$ by solving the linear program (LP)
\begin{equation}\label{eq:timesharing}
    \begin{aligned}
        \text{find}\;\; & \boldsymbol{\alpha}\geq\mathbf{0}                                                 \\
        \text{s.t.}\;\; & \textstyle\sum_k\alpha_k=1,\quad \sum_k\alpha_k\mathbf{b}_k\geq\mathbf{b}_{\min},
    \end{aligned}
\end{equation}
where $\mathbf{b}_k$ is the rate vector achieved under decoding order~$k$. Algorithm~\ref{alg:minpmac} outlines the procedure. The output flag indicates single-order feasibility ($\textsc{flag}=1$), time-sharing required ($\textsc{flag}=2$), or infeasibility ($\textsc{flag}=0$).

\begin{algorithm}[h]
    \DontPrintSemicolon
    \KwIn{$\mathbf{H}$, $\mathbf{b}_{\min}$, $\mathbf{w}$, $c_b$}
    \KwOut{$\textsc{flag}$, $\{b_u\}$, $\{\mathbf{R}_{u,n}\}$, fractions $\boldsymbol{\alpha}$}
    Initialize ellipsoid $(\mathbf{A},\boldsymbol{\theta})$\;
    \For{$\mathrm{iter}=1,2,\ldots$}{
        \ForPar{$n=1,\ldots,\bar{N}$}{
            Solve per-tone Lagrangian via L-BFGS\;
        }
        $g_u\leftarrow b_u^\star - b_{u,\min}$ for all $u$\;
        \lIf{$\sqrt{\mathbf{g}^T\mathbf{A}\mathbf{g}}\leq\epsilon$}{\textbf{break}}
        Update $(\boldsymbol{\theta},\mathbf{A})$ via ellipsoid; project $\boldsymbol{\theta}\geq\mathbf{0}$\;
    }
    Cluster users by equal $\theta_u$ values\;
    Enumerate orderings; compute $\mathbf{b}_k$ per order via~\eqref{eq:sic_rate}\;
    \eIf{single order achieves $\mathbf{b}_{\min}$}{$\textsc{flag}\leftarrow 1$}{
        Solve LP~\eqref{eq:timesharing} for $\boldsymbol{\alpha}$; $\textsc{flag}\leftarrow 2$ (or $0$ if infeasible)\;
    }
    \caption{\textsc{minPMAC}}\label{alg:minpmac}
\end{algorithm}

\subsection{{\upshape\textsc{maxRESMAC}}: Energy-constrained rate maximization}\label{sec:maxresmac}
When a single total energy constraint $\sum_u\mathcal{E}_u\leq\mathcal{E}_{T}$ replaces the per-user budgets, the dual collapses to a single scalar multiplier~$\lambda$,
\begin{equation}\label{eq:maxresmac}
    \begin{aligned}
        \maximize_{\{\mathbf{R}_{u,n}\succeq 0\}} & \;\; \sum_{u=1}^{U}\theta_u\, b_u                                                                \\
        \text{s.t.}                               & \;\; \sum_{u=1}^{U}\sum_{n=1}^{\bar{N}}\mathrm{tr}\big(\mathbf{R}_{u,n}\big)\leq\mathcal{E}_{T}.
    \end{aligned}
\end{equation}
The per-tone Lagrangian retains the form~\eqref{eq:tone_lag} with a common $\lambda$ replacing every per-user~$w_u$. Because total energy decreases monotonically in~$\lambda$, the optimal $\lambda$ is found by \emph{bisection} on the geometric mean $\lambda\leftarrow\sqrt{\lambda_{\min}\lambda_{\max}}$, with each step solving all $\bar{N}$ tones in parallel via L-BFGS. After convergence, a proportional scaling $\mathbf{R}_{u,n}\leftarrow(\mathcal{E}_{T}/\hat{\mathcal{E}}_{T})\mathbf{R}_{u,n}$ ensures exact constraint satisfaction. This scalar structure reduces the outer loop from the ellipsoid method's $\mathcal{O}(U^2\log(1/\epsilon))$ iterations to $\mathcal{O}(\log(1/\epsilon))$ bisection steps.

\subsection{{\upshape\textsc{admMAC}}: Rate-region feasibility test}\label{sec:admmac}
Given a target rate vector $\mathbf{b}$ and per-user energies $\boldsymbol{\mathcal{E}}$, \textsc{admMAC} determines whether $\mathbf{b}\!\in\!\mathcal{C}(\boldsymbol{\mathcal{E}})$ by iteratively calling \textsc{maxRMAC} with adaptively updated weights~$\boldsymbol{\theta}$. Each iteration proceeds as follows: \textsc{maxRMAC} returns a boundary vertex~$\mathbf{b}_v$ for the current~$\boldsymbol{\theta}$, and the gap $\mathbf{g}=\mathbf{b}_v-\mathbf{b}$ is computed. Three outcomes are possible. If $\boldsymbol{\theta}^T\mathbf{g}<0$, the weighted sum-rate at the boundary already falls below $\boldsymbol{\theta}^T\mathbf{b}$, so a separating hyperplane certifies infeasibility. If $\mathbf{g}\!\geq\!\mathbf{0}$ component-wise, a single decoding order already achieves the target. Otherwise, $\boldsymbol{\theta}$ is updated to steer the next vertex toward the target by increasing $\theta_u$ for deficit users.

When users share equal $\theta$ values (clusters), the solver collects vertices from different within-cluster orderings and applies a \emph{Frank-Wolfe} convex-hull membership test to determine whether the target lies inside their convex hull. The test operates per cluster, since users in different clusters decouple when inter-cluster $\theta$ values are distinct. If the target is found inside the hull, time-sharing fractions are computed. Algorithm~\ref{alg:admmac} summarizes the procedure.

\begin{algorithm}[h]
    \DontPrintSemicolon
    \KwIn{$\mathbf{H}$, $\mathbf{b}$, $\boldsymbol{\mathcal{E}}$, $c_b$}
    \KwOut{$\textsc{flag}\in\{0,1,2\}$, $\{b_u\}$, vertex info}
    Initialize $\boldsymbol{\theta}$; $\mathcal{V}\leftarrow\emptyset$\;
    \For{$\mathrm{iter}=1,2,\ldots$}{
        $(\hat{\mathbf{R}}_{u,n},\mathbf{w},\mathbf{b}_v)\leftarrow\textsc{maxRMAC}(\mathbf{H},\boldsymbol{\mathcal{E}},\boldsymbol{\theta},c_b)$\;
        $\mathbf{g}\leftarrow\mathbf{b}_v-\mathbf{b}$\;
        \lIf{$\boldsymbol{\theta}^T\mathbf{g}<-\epsilon$}{\Return $\textsc{flag}=0$ \tcp*[f]{separating hyperplane found}}
        \lIf{$\min_u g_u\geq 0$}{\Return $\textsc{flag}=1$, single order}
        Identify clusters from equal $\theta_u$ values\;
        \ForEach{cluster $\mathcal{S}_c$}{
            Enumerate within-cluster orderings; compute vertices\;
            $\mathcal{V}\leftarrow\mathcal{V}\cup\{\text{new vertices}\}$\;
            \If{Frank-Wolfe confirms $\mathbf{b}_{\mathcal{S}_c}\in\mathrm{conv}(\mathcal{V})$}{
                Compute time-sharing fractions\;
                \Return $\textsc{flag}=2$\;
            }
        }
        Update $\boldsymbol{\theta}$; increase $\theta_u$ for users with $g_u<0$\;
    }
    \caption{\textsc{admMAC}}\label{alg:admmac}
\end{algorithm}

\subsection{Duality and interplay}\label{sec:duality}
The four solvers connect through Lagrangian duality and the geometry of the MAC capacity region (Fig.~\ref{fig:solver_relationships}). Problems~\eqref{eq:maxrmac} and~\eqref{eq:minpmac} are formal Lagrangian duals. At optimality, the KKT conditions yield complementary slackness $w_u^\star(\mathcal{E}_u - \hat{\mathcal{E}}_u)=0$ and $\theta_u^\star(b_u^\star-b_{u,\min})=0$, and strong duality holds by Slater's condition~\cite{boyd2004convex} since any strictly positive-definite covariance assignment is strictly feasible for both problems. \textsc{maxRESMAC} specializes \textsc{maxRMAC} to a single sum-energy constraint, and \textsc{admMAC} wraps \textsc{maxRMAC} in an outer feasibility loop.

Together, these solvers span both the rate and energy perspectives of the capacity region. The capacity region $\mathcal{C}(\boldsymbol{\mathcal{E}})$ collects all rate vectors achievable under per-user energy constraints $\boldsymbol{\mathcal{E}}$, and \textsc{admMAC} tests membership in it. Dually, the energy region $\mathcal{C}'(\mathbf{b})$ collects all per-user energy vectors sufficient to achieve at least rate $\mathbf{b}$, and \textsc{minPMAC} finds its minimum weighted-energy point. The boundary of $\mathcal{C}(\boldsymbol{\mathcal{E}})$ can be traced by invoking \textsc{admMAC} with gradually increasing targets until infeasibility is detected.

Finally, through \emph{BC-MAC duality}~\cite{1291726}, every BC design reduces to a dual MAC problem. The dual BC channel is formed by conjugate-transposing each user's channel matrix with a reversed user ordering, and optimal BC transmit covariances are recovered from the MAC covariances via a recursive SVD-based transformation~\cite{1327794} that equates per-user mutual information across the dual pair. This link enables canonical BC precoder design directly through the solvers above.

\section{Experimental Results}\label{sec:results}

\subsection{Experimental setup}\label{sec:setup}
All experiments consider a MIMO orthogonal frequency-division multiplexing (OFDM) uplink with $U$ transmitters sharing a common $L_y$-antenna receiver across $\bar{N}$ frequency tones in complex baseband. Unless otherwise stated, the default configuration uses $U\!=\!4$, $L_y\!=\!4$, $L_{x,u}\!=\!2$, and $\bar{N}\!=\!16$, with channels drawn from a spatially correlated Rayleigh fading model consistent with the 3GPP TR~38.901 UMi street-canyon scenario~\cite{zhu20213gpp}. Results are averaged over 100 independent channel realizations. All solvers are implemented in MATLAB with a C++ compiled backend and run on an Intel Core Ultra 9 285K with 96\,GB of memory.

Six methods are compared. (1)~The proposed solvers. (2)~MOSEK, a commercial convex solver used as the optimal reference. (3)~WMMSE with SIC~\cite{5199574}. (4)~SCA with SIC~\cite{1665016}. (5)~NOMA, a rank-1 SIC heuristic with dominant-eigenmode beamforming. (6)~Orthogonal multiple access (OMA) with greedy tone assignment and per-user waterfilling.

\begin{figure}[t]
    \centering
    \includegraphics[width=\linewidth]{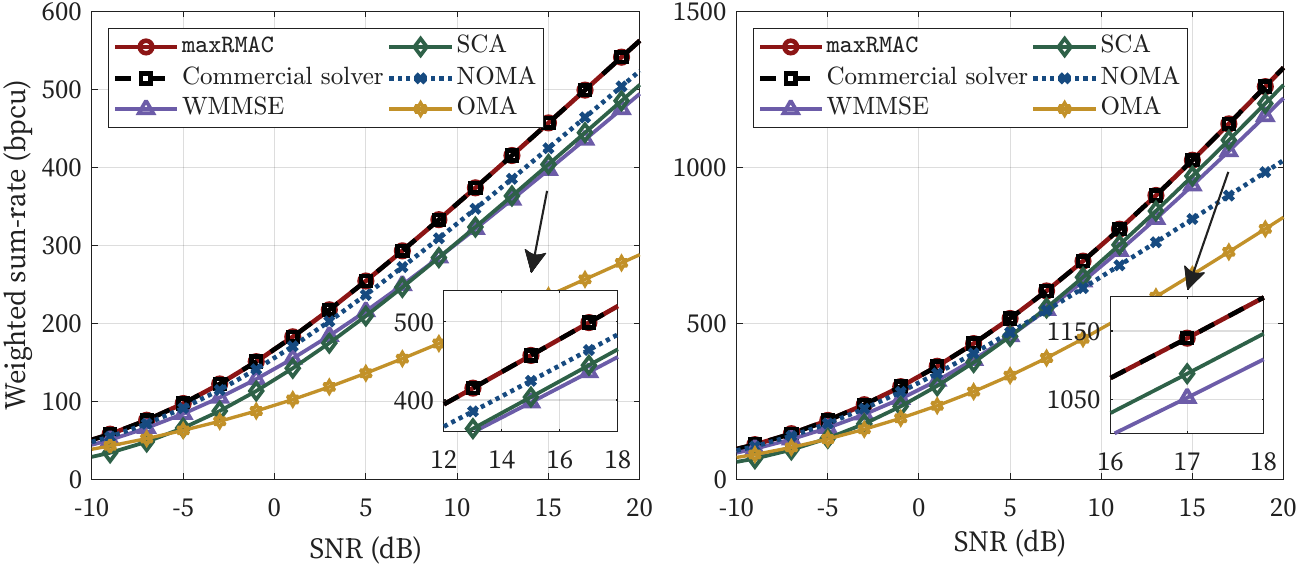}
    \caption{\textbf{Weighted sum-rate vs.\ SNR.} (left) Equal weights $\boldsymbol{\theta}\!=\![1,1,1,1]$. (right) Asymmetric weights $\boldsymbol{\theta}\!=\![4,2,1,0.5]$. \textsc{maxRMAC} coincides with the commercial solver across the entire range (see insets), confirming global optimality.}
    \label{fig:maxrmac}
\end{figure}
\begin{figure}[t]
    \centering
    \includegraphics[width=\linewidth]{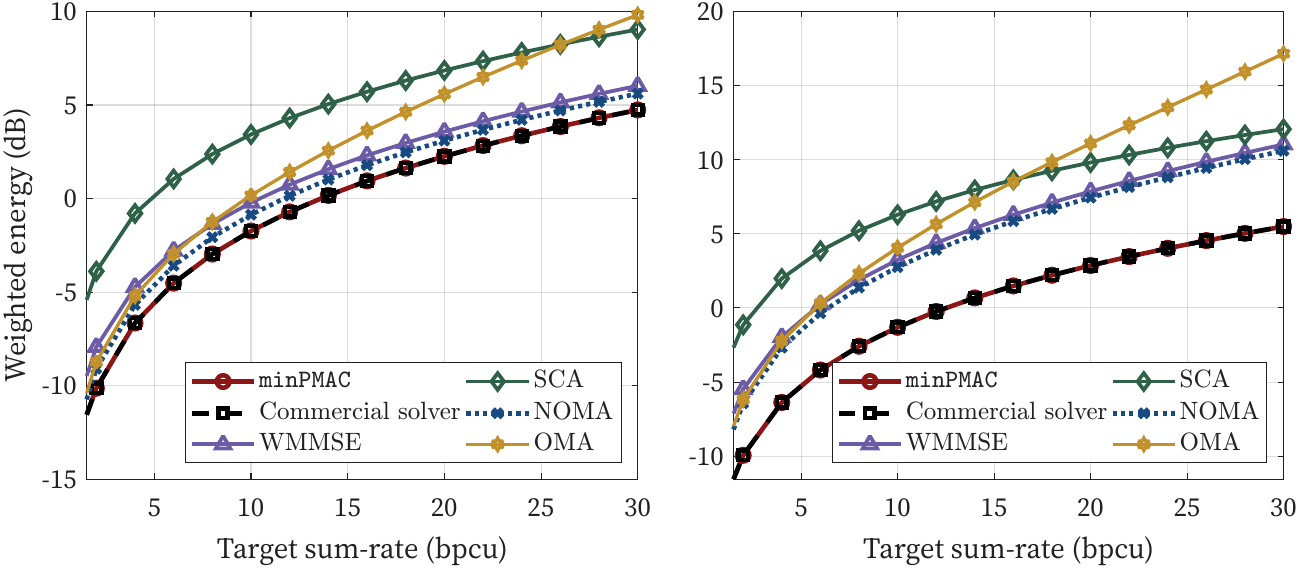}
    \caption{\textbf{Minimum weighted energy vs.\ target sum-rate.} (left) Equal per-user rate allocation. (right) Asymmetric allocation $b_u\!\propto\![4,2,1,0.5]$. \textsc{minPMAC} matches the commercial solver; all baselines require strictly more energy.}
    \label{fig:minpmac}
\end{figure}
\begin{figure}[t]
    \centering
    \includegraphics[width=\linewidth]{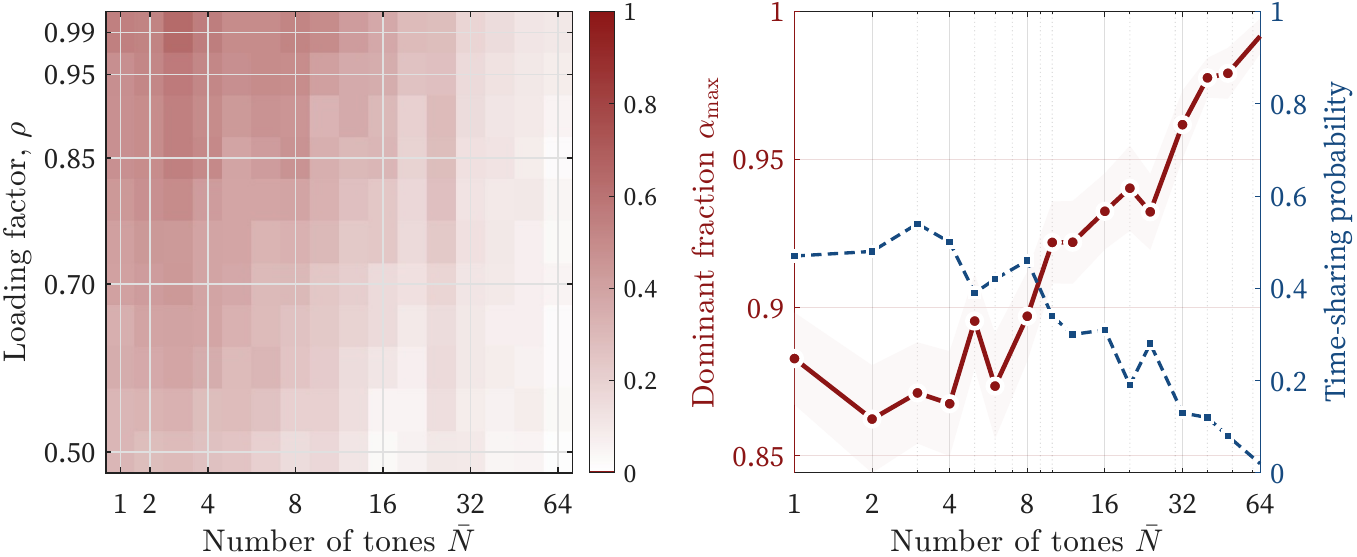}
    \caption{\textbf{Time-sharing structure} ($U\!=\!3$, $L_y\!=\!2$, $L_{x,u}\!=\!1$, frequency-selective channel). (left) Time-sharing probability vs.\ $\bar{N}$ and loading factor $\rho$. (right) Dominant fraction $\alpha_{\max}$ (left axis) and time-sharing probability (right axis) at $\rho\!=\!0.85$. Time-sharing vanishes as $\bar{N}$ grows.}
    \label{fig:time_sharing}
\end{figure}
\begin{figure*}[t]
    \centering
    \includegraphics[width=\linewidth]{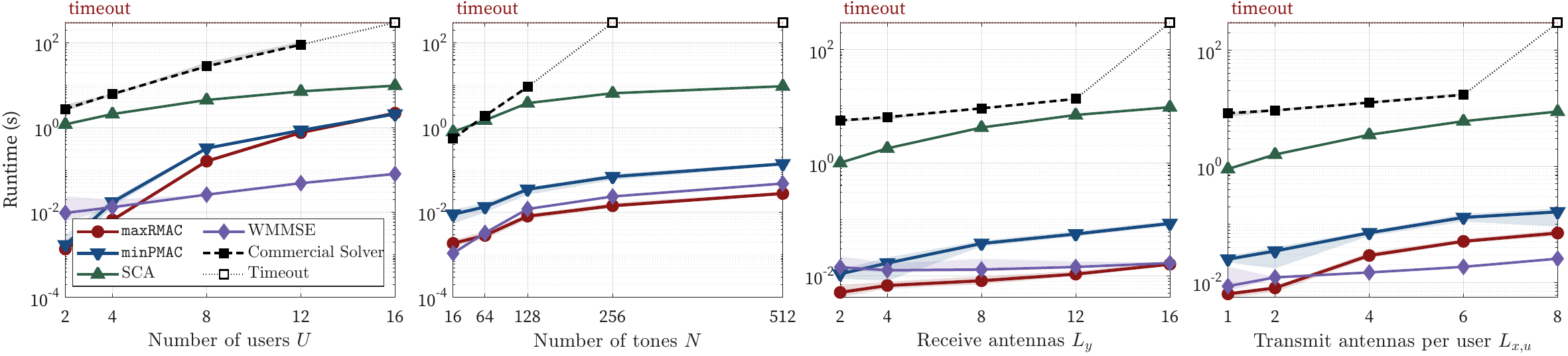}
    \caption{\textbf{Runtime vs.\ problem size} at $\mathrm{SNR}\!=\!15$\,dB with i.i.d.\ Rayleigh channels. Sweeps over $U$, $\bar{N}$, $L_y$, and $L_{x,u}$; open markers denote timeouts. The proposed solvers are 10 to 100$\times$ faster than the commercial solver across all dimensions.}
    \label{fig:scalability}
\end{figure*}

\subsection{Results}\label{sec:results_sub}
\textbf{Weighted sum-rate maximization.}
Fig.~\ref{fig:maxrmac} plots weighted sum-rate versus signal-to-noise ratio (SNR) from $-10$ to $20$\,dB under equal weights (left) and asymmetric weights $\boldsymbol{\theta}\!=\![4,2,1,0.5]$ (right), with per-user energy $\mathcal{E}_u\!=\!\bar{N}\!\cdot\!\mathrm{SNR}_{\mathrm{lin}}$. \textsc{maxRMAC} is indistinguishable from the commercial solver in both settings, confirming global optimality. The insets show that WMMSE and SCA saturate below the optimum at high SNR, with a gap that widens under asymmetric weights as the non-convexity of their local-search formulations becomes more pronounced. NOMA is further penalized by its rank-1 constraint, and OMA consistently performs worst due to its orthogonal channel partitioning.

\textbf{Minimum energy for target rates.}
Fig.~\ref{fig:minpmac} presents the dual picture, plotting minimum weighted energy versus a prescribed per-user rate vector swept from 2 to 30 bits per channel use (bpcu) in total. \textsc{minPMAC} matches the commercial solver exactly under both equal and asymmetric rate profiles. All baselines consume strictly more energy, with the gap widening at higher targets. The penalty is most severe under asymmetric allocation, where suboptimal decoding-order selection in the baselines wastes energy on interference that the optimal SIC order would cancel.

\textbf{Time-sharing structure.}
Fig.~\ref{fig:time_sharing} examines when \textsc{minPMAC} requires time-sharing among multiple decoding orders. The setup uses $U\!=\!3$, $L_y\!=\!2$, $L_{x,u}\!=\!1$ (an overloaded regime where $U\!>\!L_y$) with a frequency-selective channel of $\nu\!=\!3$ taps. The left panel maps time-sharing probability over tone count $\bar{N}\!\in\!\{1,\ldots,64\}$ and loading factor $\rho\!\in\![0.50,0.99]$, where $\rho$ sets each user's target rate to a fraction~$\rho$ of its single-user boundary rate. Time-sharing concentrates at low~$\bar{N}$ and high~$\rho$, where limited frequency-domain degrees of freedom prevent any single decoding order from meeting all rate targets simultaneously. As $\bar{N}$ grows, the additional tones provide sufficient spatial-frequency diversity and time-sharing vanishes. The right panel shows that even when time-sharing occurs, the dominant fraction $\alpha_{\max}$ exceeds 0.85, so a single order within the time-share carries the vast majority of the traffic.

\textbf{Scalability.}
Fig.~\ref{fig:scalability} reports wall-clock runtime in log scale across four sweeps over $U$, $\bar{N}$, $L_y$, and $L_{x,u}$, all at $\mathrm{SNR}\!=\!15$\,dB with i.i.d.\ Rayleigh channels. The proposed solvers are one to two orders of magnitude faster than the commercial solver in every sweep. In the $\bar{N}$-sweep, runtime scales linearly for \textsc{maxRMAC} and \textsc{minPMAC} because the per-tone subproblems are solved in parallel, whereas the commercial solver's semidefinite program grows superlinearly and times out at $\bar{N}\!=\!256$. In the $U$-sweep, the commercial solver times out at $U\!=\!16$, while the proposed solvers remain around 1\,s. The per-iteration cost of the L-BFGS inner loop is $\mathcal{O}(\bar{N}(U L_x^2 L_y^2 + U^2 L_x^4))$, multiplied by $\mathcal{O}(U^2\log(1/\epsilon))$ ellipsoid iterations for the outer loop. SCA and WMMSE exhibit intermediate runtimes but sacrifice solution quality, as shown in Fig.~\ref{fig:maxrmac} and Fig.~\ref{fig:minpmac}.

\begin{figure}[t]
    \centering
    \includegraphics[width=0.71\linewidth]{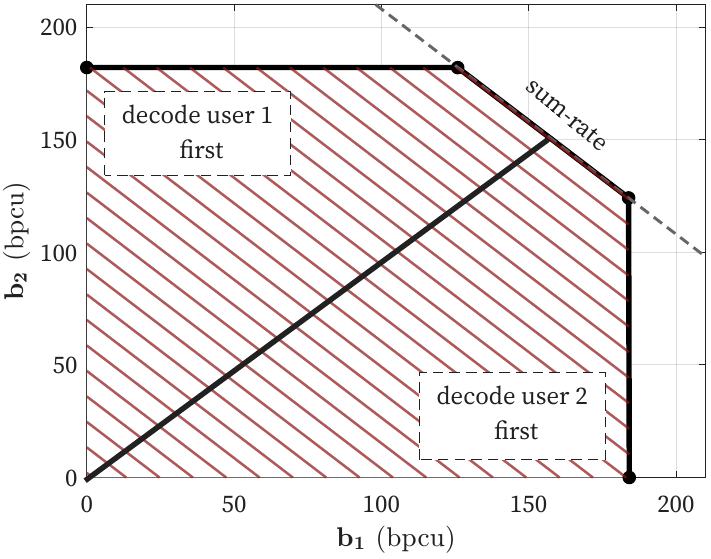}
    \caption{\textbf{MAC capacity region traced by \textsc{admMAC}} for a two-user channel with $L_y=4$, $L_{x,u}=2$, $\bar N=16$, and $\mathrm{SNR}=15$\,dB. The two corner points correspond to the two SIC orderings, and the intervening face is recovered by repeated feasibility tests.}
    \label{fig:admmac}
\end{figure}

\textbf{Capacity region tracing.}
Fig.~\ref{fig:admmac} demonstrates \textsc{admMAC} on a 2-user channel, with $L_y\!=\!4$, $L_{x,u}\!=\!2$, $\bar{N}\!=\!16$, and SNR$\!=\!15$\,dB. The solver traces the MAC capacity region boundary by bisecting on $b_2$ for each fixed $b_1$, using \textsc{admMAC} as the feasibility oracle. The two SIC corner vertices are marked, and the sum-rate line identifies the point of maximum $b_1\!+\!b_2$. The hatched feasible region is recovered from 81 boundary points, and the boundary transitions smoothly between the two vertices, with the non-trivial face achieved by time-sharing.

\section*{Acknowledgment}
This work was supported in part by Samsung START and Ericsson. The formulations and algorithms in this work build on the multiuser communication theory developed in Chapters~2 and~5 of John M. Cioffi's textbook, \textit{Data Transmission Theory}~\cite{cioffi_book}, available at \href{https://cioffi-group.stanford.edu/}{\textcolor{cardinal}{\texttt{cioffi-group.stanford.edu}}}.

\section{Conclusion}\label{sec:conclusion}
This work challenges the prevailing assumption that MIMO MAC resource allocation requires computationally expensive commercial software or suboptimal heuristic approximations. The four canonical solvers presented here achieve globally optimal capacity characterization at a fraction of the computational cost of commercial solvers and naturally extend to optimal broadcast channel precoder design through BC-MAC duality. These results establish that the MIMO MAC remains a fully tractable convex problem and that canonical solvers, rather than black-box machine learning or local-search heuristics, should serve as the standard algorithmic foundation for multiuser MIMO resource allocation.

\bibliographystyle{ieeetr}
\bibliography{ref}
\end{document}